\input xy
\xyoption{all}

\documentclass[11pt]{article}
\input epsf.tex
\usepackage{pstricks}
\usepackage{psfrag}
\usepackage{graphicx}
\usepackage[small,loose]{subfigure}
\usepackage[latin2]{inputenc}
\usepackage{epic}
\usepackage{latexsym}
\usepackage{mathrsfs}
\usepackage{bbm} % BlackBoeard letters
\usepackage{cancel} 
\usepackage{amssymb,amsmath}
\usepackage{fleqn} 
\def\harr#1#2{\smash{\mathop{\hbox to .3in{\rightarrowfill}}
 \limits^{\scriptstyle#1}_{\scriptstyle#2}}}

\def\yzero{\smash{\hbox{$y\kern-4pt\raise1pt\hbox{${}^\circ$}$}}}

\def\ov{\overline}
\def\sin2{\frac{1}{\sqrt2}}

\def\be{\begin{equation}}
\def\ee{\end{equation}}
\def\beqa{\begin{eqnarray}}
\def\eeqa{\end{eqnarray}}

\def\Dsl{\,\raise.15ex\hbox{/}\mkern-13.5mu D} %can be subscripted

\def\bfr{\begin{flushright}}
\def\efr{\end{flushright}}

\def\H{\text{H}}
\def\W{{\cal W}}

\def\IZ{\mathbb{Z}}

\def\*{\ast_{(6+1)}}

\topmargin
-.5cm
\textwidth
15.5cm
\textheight
23cm
\oddsidemargin
0.7cm
\evensidemargin
1.2cm

\begin{document}
%----------------------------------------------------------------------%
%  numbering equations with section number
%----------------------------------------------------------------------%
\makeatletter
\@addtoreset{equation}{section}
\makeatother
\renewcommand{\theequation}{\thesection.\arabic{equation}}

%-------------------------------------------------------------------------------------
%                               Title page
%_____________________________________________________________________________________

%\pagestyle{empty}

\rightline{\tt hep-th/0612088 }
\vspace{.5cm}
\begin{center}
\Large{\bf Freed-Witten anomaly in general flux compactification}\\
\vspace{1cm}

\large
Oscar Loaiza-Brito\footnote{e-mail :{\tt oloaiza@fis.cinvestav.mx}}\\[2mm] %and Tasinato$^{c,}$\footnote{e-mail: {\tt tasinato@thphys.ox.ac.uk}}\\[2mm]

{\small \em Centro de Investigaci\'on y de Estudios Avanzados, Unidad Monterrey}\\
{\small \em Cerro de las Mitras 2565, Col. Obispado, 64060, Monterrey, N.L., Mexico}\\[4mm]
{\small \em Physikalisches Institut der Universit\"at Bonn}\\
{\small \em Nussallee 12, 53115, Bonn, Germany}\\[4mm]

%\bigskip

%\large
%\\[2mm]

%$^c${\small \em The Rudolf Peierls Centre for Theoretical Physics}\\
%{\small \em Oxford University, Oxford OX1 3NP, UK}\\[4mm]

\vspace*{2cm}
\small{\bf Abstract} \\
\end{center}

\begin{center} 
\begin{minipage}[h]{14.0cm} {
Turning on a NS-NS three-form flux in a compact space drives some D-branes to be either Freed-Witten anomalous or unstable to decay into fluxes by the appearance of instantonic branes. By applying T-duality on a toroidal compactification, the NS-flux is transformed into metric fluxes. We propose a T-dual version of the Atiyah-Hirzebruch Spectral Sequence upon which we describe the Freed-Witten anomaly and the brane-flux transition driven by NS and metric fluxes in a twisted torus. The required conditions to cancel the anomaly and the appearance of new instantonic branes are also described. In addition, we give an example in which all D6-branes wrapping Freed-Witten anomaly-free three-cycles in $\widetilde{T}^6/\IZ_2\times \IZ_2$ are nevertheless unstable to be transformed into fluxes. Evenmore we find a topological transformation between RR, NS-NS and metric fluxes driven by a chain of instantonic branes.
}
\end{minipage} 
\end{center}

\bigskip

\bigskip
 
\bigskip
 
%\leftline{CINVESTAV-MTY/06-10}
%\leftline{\tt hep-th/yymmnnn}

%\leftline{May 2007}

\newpage

%--------------------------------------------------------------------------------------
%                       Paper begins
%-------------------------------------------------------------------------------------

\section{Introduction}
Incorporation of fluxes on superstring compactification setups has become an essential part of models in the literature (for a review and references therein, see \cite{Grana:2005jc}). Basically the RR and NS-NS fluxes provide a mechanism to stabilize some of the moduli. A more promising way to stabilize all of them is to compactify a type II string theory on twisted tori, or in general on nilpotent manifolds \cite{Scherk:1979zr, Kaloper:1999yr}.

Turning on external fluxes, in particular NS-NS field strengths, triggers also the appearance of other effects. For instance a D-brane cannot wrap a submanifold which in turn supports some units of NS-NS three-form flux since the system is anomalous as studied by D. Freed and E. Witten (FW) in \cite{Freed:1999vc}. In order to cancel the anomaly, one must either wrap D-branes on cycles which do not support NS-flux or to add extra branes to the anomalous ones as in Refs. \cite{Hanany:1996ie, Witten:1998xy}.
So far in the literature, many flux compactification setups have been built up by vanishing the external NS-flux on D-branes which in turn makes them FW anomaly-free.

However this is not the only one consequence of fluxes. Eventhough one avoids D-branes wrapping FW anomalous cycles, the presence of  NS-flux drives some D-branes to be unstable to decay into fluxes by the appearance of instantonic branes, as it was described by Maldacena, Moore and Seiberg (MMS) \cite{Maldacena:2001xj} (see also \cite{Diaconescu:2000wy}). This implies that the set of stable D-branes are those which are neither FW anomalous nor unstable to decay into fluxes by encountering an instantonic brane. Mathematically this establishes a connection between cohomology and twisted K-theory called the Atiyah-Hirzebruch Spectral Sequence (AHSS) (for a recent review see \cite{Evslin:2006cj} and references therein).

An interesting situation arises by compactifying type II string theory on twisted tori. Since these manifolds are realized by
taking T-duality (of simple tori) on the coordinates where a NS-flux is supported, all the information about the NS-flux in the original set up is now encoded in the metric of the twisted torus. It is said that the new torus has metric fluxes. However we know that a NS-flux imposes a stringent topological constraint, namely that some D-branes are actually stable up to the encountering with an instantonic brane. A natural question arises: what is the T-dual version of the instantonic branes in the twisted tori?

The importance of that question lies on the fact that any phenomenological model which attempts to stabilize moduli by considering extra fluxes, must be FW anomaly free and evenmore, must take care of possible topological transitions between branes and fluxes as the one described above.

Then the goal of this paper is to describe the effects of brane-flux transformation driven by instantonic branes for the case of a twisted tori compactification. As expected, we find that metric fluxes are also playing a role for driving a D-brane to be unstable. We also consider the presence of external NS-flux extra to the twisted metric and interesting enough, we find new topological transitions between RR and NS fluxes. In addition, by analyzing a particular flux configuration on a twisted six-torus on which the orbifold group $\IZ_2\times\IZ_2$ is acting, we find that all D6-branes wrapping three-cycles are unstable to be transformed into fluxes.

Our paper is organized as follows: In section 2 we give a brief review on instantonic branes in the context of the Atiyah-Hirzebruch Spectral Sequence and non-conservation of D-brane currents. In section 3 we give our proposal on the T-dual version of the AHSS. We also describe the instantonic branes for the case we have metric and NS- fluxes and some particular examples are given. Finally, we give our conclusions and some comments in the last section. Appendix A is devoted to review the AHSS.

\section{Instantonic branes}

In this section we shall briefly review some topological constraints imposed by the presence of a NS-NS flux. Essentially we shall see how the NS-NS flux drives a topological transformation between D-branes and RR and NS-NS fluxes.

\subsection{The Atiyah-Hirzebruch Spectral Sequence}

In general, a D-brane can wrap any submanifold representing a non-trivial cycle in the space homology, if such submanifold is spin$^c$ and the pullback of the NS-NS flux on the D-brane's worldvolume is trivial\footnote{There are exceptions in the context of string theory in which a cycle is not represented by  a smooth submanifold. An example of this is a seven-dimensional submanifold of the spacetime which affects a D6-brane wrapping on it. In this paper we will not consider such a case. For more details on these issues see Ref.(\cite{Evslin:2006tc}).}. In case there is  a non-trivial NS-NS flux a D-brane cannot wrap the cycle which supports it due to Freed-Witten (FW) anomaly, which makes the system inconsistent \cite{Freed:1999vc}. 

Turning on a NS-NS (NS for short) flux in type II theories restricts the number of submanifolds D-branes can wrap. Roughly speaking, this means that homology cycles are represented by a bigger set of submanifolds than those on which actually D-branes can be wrapped. Since a D-brane is charged by a RR field which in turn couples to the correspondent cycle, one expects that D-brane charges must be classified by a more refined group than (co)homology. This group turns out to be K-theory \cite{Minasian:1997mm, Witten:1998cd, Schwarz:1999vu, Witten:2000cn, Moore:2003vf}, or more explicity since we are dealing with a non-trivial NS flux, twisted K-theory (see for instance \cite{Evslin:2006cj} and references therein).

Hence it is interesting to describe how some integral cohomology classes are projected out to refine the original set into a twisted K-theory group. Physically this process was studied by Maldacena, Moore and Seiberg (MMS)in Ref.\cite{Maldacena:2001xj} and it was based on the mathematical algorithm called the Atiyah-Hirzebruch Spectral Sequence (AHSS). Interesting enough is to elucidate that one of the more fruitful objects, from the point of view of string theory, are precisely the branes wrapping those cohomology cycles which are not lifted to twisted K-theory as we shall see in due course.\\

Let us consider a D$p$-brane wrapping a spatial cycle $\Sigma_p$ in a compact manifold\footnote{Through all this paper, we consider only spin$^c$ cycles.} of dimension $n$. Applying Poincar\'e duality to it, one gets a $(n-p)$-form $\sigma_{n-p}$. Following MMS the first condition for the form to be represented by an stable D$p$-brane is that $\sigma_{n-p}$ be non-trivial in $\H^{n-p}(X;\IZ)$. The second condition is that the D-brane wrapping $\Sigma_p$ be a FW anomaly-free system which means that restricted to the D$p$ worldvolume, $H_3=0$ in cohomology. This also implies that
\begin{align}
d_3(\sigma_{n-p}):= \sigma_{n-p}\wedge H_3=0.
\end{align}
Hence, a D$p$-brane represents a form which is not only closed by the standard differential map $d$ but also by $d_3$ (notice that $d_3$ is nilpotent as well). A  D-brane representing an exact form under $d$ would wrap a cycle with boundary, which in turn implies that RR gauge invariance is violated. If one defines the Poincar\'e dual of the differential map $d_3$ by a differential operator $\partial_3$ in the sense of \cite{Collinucci:2006ug}, the straightforward question is whether a brane wrapping a cycle with boundary with respect to $\partial_3$ would be stable or not. The answer is clearly not since such a brane suffers from a  FW anomaly.\\

On the other hand, in the absence of a NS-NS flux, an stable D-brane is not only represented by a closed form but by a non-exact one as well, since otherwise the RR charge would be zero. Integral cohomology is clearly the correct group to classify (forms represented by) D-branes in a background without extra fluxes and fixed points (under some discrete group actions as orientifolds). However, as soon as the extra NS-NS flux is turned on one must only select closed forms under $d_3$. The interesting point is to elucidate what happens with the RR charge for branes representing  $d_3$-exact forms.

This was studied by MMS in \cite{Maldacena:2001xj} and one can briefly describe it as follows. Let $\sigma_{n-p}$ be the form represented by a D$p$-brane; provided $d_3\sigma_{n-p}=d\sigma_{n-p}=0$ the D$p$-brane is in principle stable. However if $\sigma_{n-p}=d_3(\eta_{n-p-3})$ the D$p$-brane would be stable, wrapping the cycle $\Sigma_p$, until it encounters a D$(p+2)$ wrapping the cycle $\Pi_{p+3}$ (Poincar\'e dual to $\eta_{n-p-3}$) in which it is immersed. The bigger D-brane must be an instantonic brane. 

The above follows by noticing that a NS flux supported in a transversal cycle to $\Sigma_p$ drives the instantonic brane anomalous. Roughly speaking, the anomaly is related to the fact that the flux $H_3$ induces a divergent magnetic field strength on the worldvolume of the instantonic brane where there are not magnetic sources. The addition of a co-dimension 3 brane provides the magnetic source which cures the anomaly. Then it is required one D-brane for each unit of NS-NS flux supported transversally to it. Physically one has that a D$p$-brane, which seems to be stable, would however disappear by encountering an instantonic D$(p+2)$-brane which supports one unit of $H_3$. 

This leads to the conclusion that stable D-branes must represent forms which belong to the group
\begin{align}
E^p_3 := \frac{\text{Ker~}d_3|_{\text{H}^p}}{\text{Im}~d_3|_{\text{H}^{p-3}}}
\end{align}
i.e., forms that are closed but not exact under $d_3$. This turns out to be the first refinement step from cohomology to twisted K-theory and for the conditions we have (a ten-dimensional spacetime, no orientifolds) it is the final step. So, we can say that those forms which survive the refinement are truly twisted K-theory classes represented by stable D-branes. This is an algorithm which connects integral cohomology to twisted K-theory and it is known as the Atiyah-Herzibruch Spectral Sequence (AHSS). A brief review of it is provided in Appendix A. Notice as well that the group $E_3^p$ can also be defined as
\begin{align}
E^p_3=\frac{\text{Ker~}d_H|_{\text{H}^p}}{\text{Im~}d_H|_{\text{H}^{p-3}}},
\end{align}
where $d_H=(d-d_3)$. This last definition will be shortly justified.\\

Now, since the instantonic brane is also magnetical charged under the RR field strength $\ast F_{p+4}$, after its disappearance the field will remain in the background and supported in transversal coordinates to those where the instantonic brane was. The transition occurs when one D$p$-brane transforms into the coupling of $\ast F_{p+4}\wedge H_3$. The charge of the disappeared D-branes is now carried by the fluxes,
\begin{align}
Q_{Dp}=\int \ast F_{p+4}\wedge H_3\;.
\end{align}
It is this process which allows to transform branes into fluxes and viceversa, which makes them to be topologically equivalent \cite{Uranga:2002vk}. 

\subsection{D-brane currents in SUGRA}
The above process, formally described by the AHSS, can be also inferred by studying the D-branes current conservation in the context of supergravity. This formalism will be helpful for the issues treated in the next section.

For the RR field strengths, we will use the democratic formulation (see e.g. \cite{Grana:2005jc} and references therein) which encodes all RR potentials,
\begin{align}
F^{(10)}=dC-H_3\wedge C,
\end{align}
where $H_3=dB_2$. The Bianchi identities for the NS flux and for the democratic RR fluxes are
\begin{align}
dH_3&=0\\
dF^{(10)}&=F^{(10)}\wedge H_3,
\end{align}
from which one sees that the Bianchi identity for the RR fields are twisted in the presence of a NS-flux. By defining a differential map 
\begin{align}
{d}_H:=(d + H_3\wedge)= d-d_3,
\end{align}
the democratic RR fields $F^{(10)}$ are indeed closed under ${d}_H$ 
\begin{align}
{d}_{H}F^{(10)}=0
\end{align}
but not under the ordinary differential $d$. 

Let us now add sources (D-branes) for the RR fluxes in the above equation,
\begin{align}
{d}_H F^{(10)}=J.
\end{align}
Notice that there are two contributions for the D-brane charge: the RR field strength $F$ and the fluxes $H_3\wedge F$. Now, since ${d}_H$ is nilpotent, ${d}_H J=0$, and the current is conserved {\it with respect to ${d}_H$}. In terms of the ordinary differential map, one gets that
\begin{align}
dJ=J\wedge H_3 =d_3J,
\end{align}
from which it is realized that the current of D-branes is not conserved due to the presence of the NS-flux $H_3$.

Therefore, the current $J$ is representing a D-brane (or in this democratic formulation, a network of D-branes) which in the AHSS context, should be instantonic. On the other hand, $dJ$ is closed under $d_3$ but exact. So it is a trivial class in $E_3$ and should represent (a network of) D-branes wrapping a non-anomalous spatial cycle which decay by encountering an instantonic brane. 

Hence supergravity equations of motion and Bianchi identities can indeed reproduce the AHSS formalism.

Let us see a more concrete case\footnote{This approach has been taken in the spirit of \cite{Townsend:1996em}. See also \cite{Evslin:2001cj} and \cite{Bekaert:2002cz}}. Consider a D$p$-brane which is charged under the RR field strength $\ast F_{p+2}$. The action involves the kinetic term for the RR fluxes and the coupling between the potential and the worldvolume. If a Chern-Simons term is present, it must be of the form
\begin{align}
\int C_{p+1}\wedge H_3\wedge \ast F_{p+4},
\end{align}
where $\ast F_{p+4}$ is another RR flux. For such a term to be non zero, the NS flux must be supported transversally to the D$p$-brane and the flux $\ast F_{p+4}$ must be transversal to both, $H_3$ and $C_{p+1}$. Being $\ast F_{p+4}$ the magnetic field strength under which a D$(p+2)$-brane is charged, a non-zero Chern-Simons term leads to a configuration of a D$p$-brane immersed in the D$(p+2)$ such that the worldvolume of the former is of codimension 3 in the worldvolume of the latter and the NS-flux being transversal to the D$p$-brane. This configuration encodes all the requirements for the MMS model to be applied. So we must somehow arrive at the same conclusion, i.e., that an instantonic brane is driving the decay of an stable D-brane, by just considering the currents. Indeed, the equation of motion for the RR potential $C_{p+1}$ is
\begin{align}
d\ast F_{p+2}= \ast J_{p+1} + H_3\wedge \ast F_{p+4}.
\end{align}
The presence of the D$(p+2)$-brane makes the current non conserved since
\begin{align}
d\ast J_{p+1}= H_3\wedge\ast J_{p+3},
\end{align}
where $\ast J_{p+3}$ is the current associated to the D$(p+2)$-brane. If this brane is instantonic, the above equation tells us that a D$p$-brane disappear by encountering it and the charge of the disappeared D$p$-brane is now carried by the fluxes $H_3\wedge \ast F_{p+4}$ in accordance with the MMS picture (see also \cite{deAlwis:2006cb}).

%=================================================================section=========================================================================
\section{Instantonic branes in twisted torus compactification}
In the past few years, IIB string compactification with extra fluxes has been useful to stabilize some moduli. One fruitful attempt to improve the above setup has been to consider compactification on its T-dual version\footnote{For a formal description  about T-duality and NS-flux, see \cite{Bouwknegt:2003vb} and \cite{Bouwknegt:2003wp}.}. By taking T-duality, the extra flux manifests itself as metric fluxes and  deforms the torus on which now type IIA is compactified. This deformed torus is called twisted torus (\cite{Scherk:1979zr, Kaloper:1999yr,Marchesano:2006ns}. 

In this section we shall study how instantonic branes are manifested in this background. Some concrete examples are given in the end of the section.

\subsection{Compactification on twisted tori}
Let us consider IIB string theory compactified on a six-dimensional flat torus $T^6$ in the presence of $N$ units of a non-trivial flux $H_3=-N\omega_3$, where 
\begin{align}
\omega_3=\sum_{i,j,k}dx^i\wedge dx^j\wedge dx^k\;.
\end{align}
After T-duality on $x^{i_1}, x^{i_2}, x^{i_3}$, the metric becomes 
\begin{align}
ds^2=\tilde{g}_{ij}\eta^i\eta^j,
\end{align}
where we denote the metric elements in the new basis $\eta$ as ${\tilde{g}_{ij}}$ and
\begin{align}
\eta^i=dx^i-\omega^i_{jk}x^jdx^k,
\end{align}
with $i$ running over 1 to 6 and the constant structure $\omega^i_{jk}=0$ for $i\in\{i_i, i_2,i_3\}$. The relation between the new and the old basis given by the above equation implies that the one-forms $\eta$ are not closed,
\begin{align}
d\eta^i=-\omega^i_{jk}\eta^j\wedge\eta^k.
\end{align}
Notice that the constant structure $\omega^i_{jk}$ would be zero had the NS-flux in the type IIB side been zero. Evenmore, it is $\omega^i_{jk}$ which makes the $\eta$'s to be non-closed.

As a consequence it is possible to define an exact two-form which has as components the structure constants $\omega^i_{jk}$. This is done through the derivative of the metric which is also non-closed,
\begin{align}
\omega_{(x)}=-dg_{(x)}=-\frac{1}{g_{xx}}\omega^x_{ab}dx^a\wedge dx^b,
\end{align}
with $x\in\{i_1,i_2,i_3\}$. The NS-flux in type IIB is now encoded in $\omega^i_{jk}$ which from now on we shall refer to as metric fluxes.
Hence, for a general $p$-form (with constant components $\sigma_{\mu_1\cdots\mu_p}$)
\begin{align}
\sigma_p=\sigma_{\mu_1\cdots\mu_p}\eta^{\mu_1}\wedge\cdots\wedge\eta^{\mu_p},
\end{align}
one gets that
\begin{align}
d\sigma_p=\omega_{(x)}\wedge\sigma_{p(x)},
\end{align}
where $\sigma_{p(x)}$ is the $(p-1)$-form defined as 
\begin{align}
\sigma_{p(x)}=\sigma_{x\mu_1\cdots\mu_{p-1}}\eta^{\mu_1}\wedge\cdots\wedge\eta^{\mu_{p-1}}. 
\end{align}
It is straightforward to show that 
\begin{align}
\omega_{(x)}\wedge \sigma_{p(x)}= -\omega^x_{ab}\sigma_{x\mu_1\cdots\mu_{p-1}}\eta^a\wedge\eta^b\wedge\eta^{\mu_1}\wedge\cdots\wedge\eta^{\mu_{p-1}}:=-\omega\sigma_p,
\end{align}
from which the differential of a $p$-form can also be written as $d\sigma_p=-\omega\sigma_p$.
Hence, in general the differential of a $p$-form will have two components: the usual derivative (for non-constant components) and $\omega\sigma_p$. This leads us to redefine the differential map as 
\begin{align}
\ov{d}_\omega = (d+\omega),
\end{align}
such that $\ov{d}_\omega \sigma_p= d\sigma_p+\omega\sigma_p$. Notice that in this notation, $\ov{d}_\omega\eta^i=0$.

Finally one sees that when $\sigma_p$ is closed under $\ov{d}_\omega$, it is not under the ordinary differential $d$. This certainly resembles the properties of ${d}_H$ in type IIB side.

\subsection{AHSS T-dual version}
We have  seen that in type IIA the metric fluxes play the role of NS-fluxes upon T-dualization on Type IIB. It is then natural to study how all topological effects driven by the NS-flux are manifested in the T-dual version by the metric fluxes. For instance, it is expected that stable D-branes represent closed forms which somehow involve the metric fluxes. The natural candidate to define which form is closed is still the differential map $d$.

Therefore, we require an algorithm which refines a set of general forms into those which are closed but not exact under $d$. Notice that by this procedure we will give ride of those forms that are closed due to the presence of the  metric fluxes $\omega_{(x)}$ which undoubtedly leads us to a small set of forms than that get it by cohomology of the flat six-torus.
This is the T-dual version of the first step in the AHSS we propose, which as we shall see, reproduces the results shown in \cite{Marchesano:2006ns}. The explicit construction is as follows.

Consider in general $p$-forms in $\Omega_p(\widetilde{T}^6)$. The first step is then to select forms that are closed but not exact under the ordinary differential map,
(which must be the T-dual version of $E^p_3(T^6)$ in type IIB), 
\begin{align}
\ov{E}^p_1=\frac{\text{Ker}~d|_{\Omega^p}}{\text{Im}~d|_{\Omega^{p-1}}},
\end{align}
which is nothing more than the $p$-th cohomology group of the twisted torus $\widetilde{T}^6$ denoted as ${\text{H}}^p(\widetilde{T}^6;\IZ)$.

Important enough is to notice that, while in type IIB with NS-flux all the topological constraints imposed by the flux were encoded in the second step in the AHSS, in its T-dual version they are contained at the level of cohomology. 

Therefore non-trivial forms in $\ov{E}^p_1(\widetilde{T}^6)$ are expected to be represented by stable D-branes. in which  the T-dual FW anomaly now reads
\begin{align}
d\sigma_p=-\omega\sigma_p=0.
\end{align}
for some $p$-form $\sigma_p$. On the other hand, 
for the twisted torus $\widetilde{T^6}$ it turns out\footnote{We shall see this in next examples.} that there are exact forms which are in fact $N$-torsion forms and henceforth they are written as $N\xi_{p+1}={d} \sigma_p=-\omega\sigma_p$. Then, for those forms, if $N\xi_{p+1}$ is exact,
\begin{align}
[N\xi_{p+1}]=[0]\in {\text{H}}^p(\widetilde{T^6};\IZ),
\end{align}
and  $[\xi_{p+1}]\in \text{Tor~H}^p(\widetilde{T^6};\IZ)$.

Upon Poincar\'e duality one can easily see that a D$p$-brane representing an exact form $\sigma_{6-p}=\ov{d}_\omega \pi_{5-p}$ is wrapping a cycle with boundary since,
\begin{align}
\Sigma_p=\partial \Pi_{p+1}
\end{align}
where $\Sigma_p=\text{PD}(\sigma_{6-p})$ and $\Pi_{p+1}=\text{PD}(\pi_{5-p})$. This is nothing more than the T-dual version of a type IIB D-brane wrapping a cycle ``with boundary'' with respect to $\partial_3$ as defined in the previous section.

\begin{figure}[t]
\begin{center}
\centering
\epsfysize=8cm
\leavevmode
\epsfbox{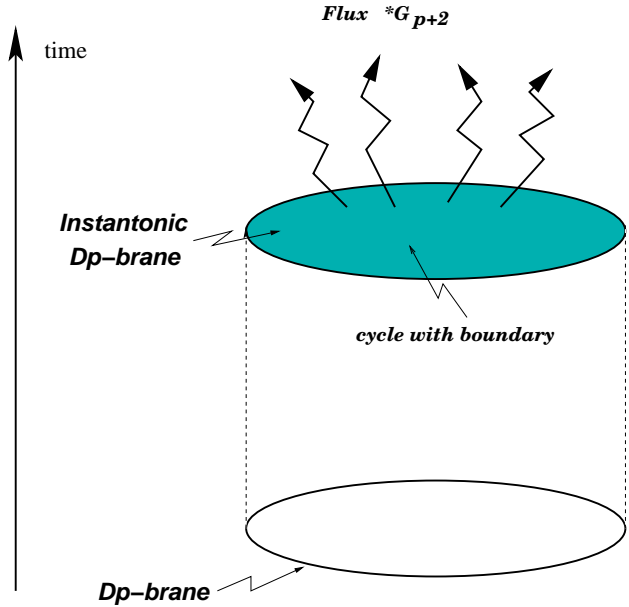}
\end{center}
\caption[]{\small In presence of metric fluxes, a Dp-brane ends at an instantonic Dp-brane. The instantonic brane emits a remanent flux $\ast G_{p+2}$.}
\label{Fig:pinst}
\end{figure}

All the above construction yields the suggestion that T-dual version of the transition between D-branes and fluxes occurs as follows: A D$p$-brane, in principle stable, wraps a cycle $\Sigma_p \in {\text{H}}_p(\widetilde{T}^6;\IZ)$ which is not torsion in the presence of $N$ units of metric flux $\omega_{(x)}$ supported transversally to $\Sigma_p$. However $N$ D$p$-brane drives the cycle to be exact since $N\Sigma_p\in \text{Tor~H}_p(\widetilde{T}^6;\IZ)$. The $N$ D$p$-brane travel in time and encounter an instantonic D$p$-brane wrapping the spatial cycle $\Pi_{p+1}$ (which supports the metric fluxes) and disappear. As a remnant the magnetic RR field strength $\ast F_{p+2}$, created by the instantonic brane, couples to metric fluxes as $-\omega\ast F_{p+2}$ from which the charge of the disappeared branes is computed 
\begin{align}
Q_{D_p}=-\int \omega\ast F_{p+2}=N.
\end{align}
The process is depicted in Fig.(\ref{Fig:pinst}).

In the following we study the above setup from the point of view of supergravity and we check consistency on the field theory on the worldvolume of a D$p$-brane supporting metric fluxes. These two cases will support the above T-dual picture of a transition between branes and fluxes.

\subsubsection{D-brane currents}
In type IIB on $T^6$ the supergravity equations of motion, together with a Chern-Simons term is in agreement with the physics of the AHSS as we have seen in the first section. Hence, one way to support the T-dual AHSS proposal is to check if it is compatible with IIA supergravity equations of  motion with a T-dual Chern-Simons term involving the presence of metric fluxes. 

The T-dual Chern-Simons term in type IIA reads \cite{Cascales:2003zp}
\begin{align}
S_{CS}=\frac{1}{2\kappa^2_0}\int_{\widetilde{T}^6}C_{p+1}\wedge\omega_{(x)}\wedge\ast F_{p+4(x)}\;.
\end{align}
The supergravity equation of motion with respect to the RR potential $C_{p+1}$ are $d\ast J_{p+1}=-\omega\ast J_{p+1}$, i.e.,
\begin{align}
\ov{d}_\omega\ast J_{p+1}=0,
\end{align}
which is compatible with a D$p$-brane action in terms of the differential $\ov{d}_\omega$, since the above CS term arises from the kinetic term,
\begin{align}
S_\text{K}&=-\frac{1}{2\kappa_0^2}\int F_{p+2}\wedge\ast F_{p+2}=\\
&=\frac{1}{2\kappa_0^2}\int C_{p+1}\wedge \ov{d}_\omega\ast F_{p+2}=\\
&=\frac{1}{2\kappa_0^2}\left(\int C_{p+1}\wedge d\ast F_{p+2}-\int C_{p+1}\wedge \omega_{(x)}\wedge \ast F_{(p+4)(x)}\right).
\end{align}
Notice that the metric fluxes are the source for the non-conservation of D$p$-brane current with respect to $d$. However, charge measured by $\ast J_{p+1}$ is indeed conserved in terms of $\ov{d}_\omega$ which means that the total contribution for the D$p$-brane charge coming from D$p$-branes and fluxes remains constant.

In the democratic formulation for RR fields, the current conservation (or non- conservation) for a D-brane network reads $\ov{d}_\omega J=0$, which is realized from the equations of motion
\begin{align}
\ov{d}_\omega {F}^{(10)}= J,
\end{align}
where ${F}^{(10)}=\ov{d}_\omega C= dC+\omega C$ as in \cite{Camara:2005dc, Villadoro:2005cu, Villadoro:2006ia}.

From these results one concludes that:
\begin{itemize}
\item
The current $J$ does not belong to ${H}(\widetilde{T}^6;\IZ)$ which for the case of the twisted torus means that
\begin{align}
dJ=N\beta
\end{align}
for some exact form $\beta$. Since the current $J$ is under Ponicar\'e duality connected to the worldvolume of a D-brane, the above equation tells us that
\begin{align}
\partial\W=N\W
\end{align}
where $\text{PD}(J)=\W$. Hence, a D-brane with current $J$ is actually wrapping a cycle with boundary. This is the T-dual of FW anomaly \cite{Marchesano:2006ns} one gets from the proposed T-dual AHSS.
\item
Since $\omega_{(x)}$ is quantized, there are $N$ units of it, from which one gets that the variation of D$p$-brane current is
\begin{align}
\int d\ast J_{p+1}=\int \omega\ast J_{p+1}=N.
\end{align}
Henceforth, the presence of metric fluxes makes the number of D$p$-branes to vary in multiples of $N$, i.e., $N$ D$p$-branes disappear if the variation is measured over time by encountering an instantonic brane.
\end{itemize}

It is followed that supergravity equations of motion and particularly Bianchi identities in type IIA compactified on a twisted torus describes the emanation or absorption of D$p$-branes by an instantonic D$p$-brane wrapped on an spatial cycle with boundary, which in turns supports a non-trivial amount of metric flux. This is totally in agreement with the T-dual AHSS we have proposed.

\subsubsection{Field theory on a D-brane}
In order to give more support to our proposal, let us study the field content on the worldvolume of a D$p$-brane which actually supports $N$ units of  metric flux. Our expectatives are to find some kind of inconsistency in similarity with the FW anomaly in type IIB with NS-flux as in \cite{Cascales:2003zp}.

Let us consider a D$p$-brane wrapping a cycle which also supports the metric flux. Hence, in the worldvolume $\W_{p+1}$ of the D$p$-brane there is a term induced by the presence of $\omega_{(x)}$,
\begin{align}
\int_{\W_{p+1}} \omega_{(x)}\wedge \widetilde{A}_{p(x)}=-\int_{\W_{p+1}}\omega\widetilde{A}_p\;.
\end{align}
Since $\omega_{(x)}$ is an exact two-form, the above term is actually zero if and only if the cycle $\W_{p+1}$ is boundaryless. However, as it was suggested by the previous case, the interesting case emerges when one considers a D-brane wrapping a cycle with boundary. So let us take the worldvolume $\W_{p+1}$ with boundary in which case the equations of motion for the induced potential $\widetilde{A}_p$ can not be fulfilled. Hence, to remove the tadpole in $\widetilde{T^6}$ let us add magnetic sources for the induced gauge potential $\widetilde{A}_p$, so that the action on $\W_{p+1}$ is
\begin{align}
\int_{\W_{p+1}}\widetilde{F}_{p+1(x)}\wedge\ast\widetilde{F}_{p+1(x)}~+~\int_{\W_{p+1}} \omega_{(x)}\wedge \widetilde{A}_{p(x)}.
\end{align}
After integrating by parts, the equations of motion are,
\begin{align}
d\ast\widetilde{F}_{p+1(x)}=\omega_{(x)}
\end{align}
which tells us that the inconsistency driven by the presence of metric fluxes is cured by adding D$p$-branes. For the initial D$p$-brane to be instantonic, one concludes that $N$ $Dp$-branes would disappeared by encountering it, leaving as a remnant the magnetic RR flux created by the instantonic brane coupled to the metric fluxes, i.e., -$\omega\ast F_{p+2}$. This agrees with the proposed the T-dual AHSS.

\subsubsection{An example}
Before continuing on constructing the next step in the T-dual AHSS, let us study a concrete example. Consider type IIB on $T^6$ with an extra NS-flux \cite{Scherk:1979zr, Kachru:2002sk, Kaloper:1999yr, Camara:2005dc, Villadoro:2006ia, Villadoro:2005cu}
\begin{align}
H_3=-N\omega_3=-N(dx^1\wedge dx^5\wedge dx^6 + dx^4\wedge dx^2\wedge dx^6 + dx^4\wedge dx^5\wedge dx^3)\;,
\end{align}
where
\begin{align}
\int_{\Sigma_3}H_3=N,
\end{align}
and $\Sigma_3$ is the Poincar\'e dual of $\omega_3$. By taking T-duality on coordinates $x^1,x^2,x^3$ we get type IIA string theory compactified on a twisted torus $\widetilde{T}^6$ with constant structures $\omega^i_{jk}$ given by
\begin{align}
\omega^1_{56}=\omega^2_{64}=\omega^3_{45}=-N\;.
\end{align}
From the (co)homology of $\widetilde{T}^6$, explicity computed in Ref.\cite{Marchesano:2006ns}, one can take branes representing $N$-torsion forms, which according to our proposal would disappear by encountering an instantonic brane representing a non-closed form. We shall denote the $p$-form $\eta^{i_1}\wedge\cdots\wedge\eta^{i_p}$ as $\eta^{i_1\cdots i_p}$ and its corresponding $(6-p)$-cycle as
\begin{align}
\text{PD}(\eta^{i_1\cdots i_p})=\Sigma_{l_1\cdots l_{6-p}} \qquad \text{with ~} \{i_1,\cdots ,i_p\}\neq \{l_1,\cdots ,l_{6-p}\}.
\end{align}
In terms of a 1-cycle basis in the twisted torus, one-cycles are expressed as
\begin{align}
\Sigma_i=(1,0)_i \qquad \text{and} \qquad \Sigma_{i+3}=(0,1)_{i+3},
\end{align}
where $i=1,2,3$ denotes in the right hand side, the two-dimensional tours under consideration. Once we have fixed our notation let us consider D6-branes wrapped on three-cycles.In particular if the three-cycle is $N$-torsion, a brane wrapping it would suffer a topological transformation. So, in order to know how many torsion three-cycles there are, we must compute the third cohomology group of the twisted torus. This is \cite{Marchesano:2006ns},
\begin{align}
\text{H}^3(\widetilde{T}^6;\IZ)=\IZ^{12}\times \IZ^4_N,
\end{align}
which tells us that there are four $N$-torsion forms and four $N$-torsion cycles. Let us select the cycle
\begin{align}
\Sigma_{123}=(1,0)_1\times(1,0)_2\times (1,0)_3,
\end{align} which is neither a non-closed nor an exact form. Hence a D6-brane wrapped there is stable. However, if we wrap not a single one but $N$ D6-branes, the cycle turns out to be the boundary of a four-cycle,
\begin{align}
\Sigma_{123}=\partial \Pi_4,
\end{align}
since now $N\Sigma_{123}\in \text{Tor~H}_3(\widetilde{T}^6;\IZ)$ and where $\Pi_4=\left[(\Sigma_{25}+\Sigma_{14})\times\Sigma_{36}-\Sigma_{1245}\right]$. Under Poincar\'e duality the correspondent forms satisfy
\begin{align}
N\eta^{456}=d(\eta^{14}-\eta^{25}-\eta^{63}):=d\zeta_2.
\end{align}
According to our proposal, $N$ D6-branes wrapping $\Sigma_{123}$ are stable until encounter an instantonic D6-brane wrapping the cycle $\Pi_4$ which in turn supports the metric flux,
\begin{align}
\omega_{(x)}=\omega^1_{45}\eta^{45}+\omega^2_{46}\eta^{46}+\omega^3_{56}\eta^{56}
\end{align}
which indeed is supported transversally to $\Sigma_{123}$. 

After the appearance of the instantonic brane, the $N$ D6-branes are transformed into flux. If our proposal is correct, the final flux must be proportional to the forms $\eta^{456}$ which was represented by the disappeared D6-branes. Indeed the magnetical RR field strength $F_2$ left as remnant by the instantonic brane is supported on the cycle
\begin{align}
\Theta_2 = \Sigma_{14}-\Sigma_{25}-\Sigma_{63}
\end{align}
which satisfies $\Theta_2\times\Pi_4=0$. Hence the RR flux is written in the $\eta$-basis as
\begin{align}
F_2=F_{14}\eta^{14}-F_{25}\eta^{25}-F_{63}\eta^{63},
\end{align}
and consequently, the coupling between this flux and the metric one is given by
\begin{align}
\omega F_2=-N\eta^{456}
\end{align}
as represented by the already disappeared D6-branes. This confirms that the T-dual AHSS works for this example.

\subsection{AHHS T-dual version and NS-flux}
Up to now we have not considered the presence of extra NS-fluxes in the twisted tori manly because we were interested in the T-dual picture of NS-flux in type IIB. The classes obtained after applying the T-dual version of the AHSS correspond to a twisted K-theory group. However from the type IIA side this refinement is taking place on the torsion part of cohomology which  comes from T-duality of the NS-flux. The cohomology classes belonging to the non-torsion part have been not refined into a K-theory class which leads us to think on a further refinement. Naively one can think that turning an extra NS-flux will do it. 

However we already have a three-form which has not been considered so far; the K\"ahler form $J$ in the twisted tori is not anymore closed under ${d}$ making that our background has a natural three-form flux $\omega J$. Hence the desired refinement from $\ov{E}_1^p(\widetilde{T}^6)$ to twisted K-theory would involve all possible three-form fluxes, this is, a NS-flux ${H_3}$ and the metric component of the K\"ahler form $\omega J$. Our proposal for a third step in the T-dual AHSS is as follows.\\

In general, by having a $B$-field one can built up a complexified K\"ahler form
\begin{align}
J_c=B+iJ.
\end{align}
In a non-twisted torus $H_3=dJ_c$ is the responsible for the apparition of the FW anomaly. In the present case
\begin{align}
\ov{d}_\omega J_c= H_3+\omega J_c.
\end{align}
This suggests that the next step in the T-dual AHHS is to take forms which are closed but not exact under the map
\begin{align}
\ov{d}_3:=(H_3+\omega J_c)\wedge\;.
\end{align}
Therefore a $p$-form represented by  a FW anomaly-free brane must satisfies
\begin{align}
\ov{d}_3~\sigma_p=(H_3+\omega J_c)\wedge \sigma_p=0,
\end{align}
which implies that at the level of cohomology
\begin{align}
H_3+\omega J_c=0.
\label{Eq:FWt}
\end{align}
Henceforth, the most direct form to cancel the FW anomaly in twisted tori is to fulfilled the above condition. Interesting enough, Eq(\ref{Eq:FWt}) has been analyzed  as the condition to assure gauge invariance in the effective theory as in \cite{Camara:2005dc} and by studying localized Bianchi identities as in \cite{Villadoro:2005cu, Villadoro:2006ia}. However for the case the three-forms fluxes are not trivial in cohomology, canceling the FW anomaly requires extra branes, which in turns will leads us to more brane-flux transitions.

Our proposal for the third refinement in the T-dual AHSS is then to consider forms which belong to the group
\begin{align}
\ov{E}^p_3(\widetilde{T}^6)~=~\frac{\text{Ker~}\ov{d}_3|_{\ov{H}^p}}{\text{Im~}\ov{d}_3|_{\ov{H}^{p-3}}}.
\end{align}

Hence, according to this, non-closed forms under ${d}_H$ cannot be lifted to a twisted K-theory class of $\widetilde{T}^6$ and D-branes representing them are physically unstable. On the other hand, exact forms are represented by D-branes which are stable until they encounter an instantonic brane (representing a closed form). After that the branes transform into a configuration of fluxes involving NS and metric ones.

Notice as well that by defining the map
\begin{align}
\ov{d}_H=d-(H_3+\omega J_c)\wedge,
\end{align}
the group of non-trivial forms under $\ov{d}_3$ can also be written as
\begin{align}
\ov{E}^p_3(\widetilde{T}^6)=\frac{\text{Ker~}\ov{d}_H|_{\ov{\text{H}}^p}}{\text{Im~}\ov{d}_H|_{\text{H}^{p-3}}}.
\end{align}
In this context one can see that  a $\ov{d}_3$ non-closed form $\sigma_p$ satisfies $d_3 \sigma_p=d\eta_{p-1}=-\omega\eta_{p-1}$ which suggests that the flux representing $\omega\eta_{p-1}$ can also be turned into D-branes. We shall return to this later on.

\subsubsection{RR fields equations of motion}
In order to support our previous statements, let us study the RR field equations of motion in the presence of metric and NS-fluxes. We shall see that the disappearance of D-branes is an straightforward consequence of having extra fluxes. Specifically the introduction of metric fluxes and a B-field\footnote{There is not an induced D-brane charge by the B-field mainly because it is canceled against bulk Chern-Simons terms as it was shown in \cite{Taylor:2000za}.} induces a Wess-Zumino term given by
\begin{align}
I_{p+4}=\mu_{p+2}\int_{\W_{p+3}}C_{p+1}\wedge (B_2+iJ).
\end{align}
Using the relation $\ast J_{p+3}=\ov{d}_\omega\ast {F}_{p+4}=-2\kappa_0^2\mu_{p+2}\text{PD}(\W_{p+3})$ one gets
\begin{align}
I_{p+4}=-\frac{1}{2\kappa_0^2}\int_{\widetilde{T}^6} C_{p+1}\wedge J_c\wedge \ov{d}_\omega\ast {F}_{p+4},
\end{align}
which is actually a Chern-Simons term. Hence the equations of motion for the RR field becomes after integration by parts,
\begin{align}
\ov{d}_\omega\ast {F}_{p+2}=\ast J_{p+1}-({H}_3+\omega J_c)\wedge\ast {F}_{p+4}.
\end{align}
The current is then non-conserved if there are D$(p+2)$-branes,
\begin{align}
\ov{d}_\omega\ast J_{p+1}=({H}_3+\omega J_c)\wedge\ast J_{p+3}.
\end{align}
By defining the differential map
\begin{align}
\ov{d}_H={d}-\ov{d}_3
\end{align}
with $\ov{d}_3=({H}_3+\omega J_c)\wedge$ one gets that the current is indeed conserved under $\ov{d}_H$,
\begin{align}
\ov{d}_H\ast J_{p+1}=0.
\end{align}

In the democratic formulation of the RR fields, the twisted Bianchi identity and the non-conservation of the current read,
\begin{align}
\ov{d}_\omega B_2&=({H}_3+\omega J_c),\\
{d}{F}^{(10)}&=\ov{d}_3{F}^{(10)}+J,\\
\ov{d}_H J&=0.
\end{align}
Then the current $\ast J_{p+1}$ does not belong to $\ov{E}_3^{5-p}(\widetilde{T}^6)$, proving that it does represent an instantonic brane. In such case a current of $N$ D$p$-branes emanates from it provided $N$ units of three-form flux ${H}_3+\omega J_c$ are transversally supported to the cycle where the D$p$-branes are wrapped.

After the disappearance of the $N$ D$p$-branes, there is a remnant of fluxes
\begin{align}
({H}_3+\omega J_c)\wedge\ast F_{p+4}
\end{align}
carrying the D$p$-brane charges.

\subsubsection{Field theory}
As in previous cases, the presence of ${H}_3+\omega J_c$ induces an action term on a D$(p+2)$-brane's worldvolume
\begin{align}
\int_{\W_{p+3}}(H_3+\omega J_c)\wedge\widetilde{A}_p,
\end{align}
driving the field theory inconsistent.

The way to cure it is to add magnetic sources for the gauge potential $\widetilde{A}_p$, after which the equation of motion becomes
\begin{align}
\ov{d}_\omega\ast\widetilde{{F}}_{p+1}={H}_3+\omega J_c.
\end{align}
In the case the D$(p+2)$-brane is instantonic, the sources can only be D$p$-branes terminating at the former one. This configuration is in agreement with the proposed T-dual AHSS.

\subsection{Example 2: Twisted six-torus threaded by a constant NS-NS flux}

Let us consider the same twisted six-torus as in Example 1. In this case the complexified K\"ahler form $J_c=B+iJ$ is \cite{Marchesano:2006ns}
\begin{align}
J_c=-\frac{1}{2}\sum_{i=1}^3\left(\frac{T_i}{\text{Im~}\tau^i}e^i\wedge\bar{e}^i\right),
\end{align}
where
\begin{align}
e^i=\eta^i+\tau^i\eta^{i+3}
\end{align}
for $i=1,2,3$ and $\tau^i$ being the complex structure parameters whereas $T_j$ are the complexified K\"ahler parameters. Therefore,
\begin{align}
H_3+\omega J_c=&\ov{d}_\omega J_c=M\left(-\frac{T_1}{\text{Im~}\tau^i}\eta^{156}+\frac{T_2}{\text{Im~}\tau^2}\eta^{264}+\frac{T_3}{\text{Im~}\tau^3}\eta^{345}\right)+\nonumber\\
&+M\left(\text{Re~}\tau^1+\text{Re~}\tau^2+\text{Re~}\tau^3\right)\eta^{456}.
\label{hfield}
\end{align}

For simplicity let us only consider for the moment the term $-\frac{T_1}{\text{Im~}\tau^1}M\eta^{156}$. It follows that a zero-form $\phi \in \text{H}^0(\widetilde{T}^6;\IZ)$ is non-closed under $\ov{d}_3$ since
\begin{align}
\ov{d}_3\phi=-M\frac{T_1}{\text{Im~}\tau^1}\eta^{156}\neq 0.
\end{align}
Hence an space-filling D6-brane representing the form $\eta^{156}$ is exact and consequently trivial in $\ov{E}_3^3(\widetilde{T}^6)$. It also wraps the cycle
\begin{align}
\text{PD}(\eta^{156})=\Sigma_{234}
\end{align}
which is non-trivial in $H_3(\widetilde{T}^6;\IZ)$.

On the other hand, the zero-form $\phi$ is representing an space-filling D8-brane wrapping the whole twisted six-torus. Therefore $\phi$ does not belong to $\ov{E}^0_3(\widetilde{T}^6)$ and it is not an stable D8-brane but rather an instantonic one.

According to the proposed T-dual AHSS, $M$ D6-branes wrapping $\Sigma_{234}$ would turn into flux by encountering an instantonic D8-brane wrapping $\widetilde{T}^6$. Its magnetic field $F_0$ couples to ${H}_3+\omega J_c$ after the D8 disappearance as
\begin{align}
F_0\wedge\ov{H}_3=-\frac{T_1}{\text{Im~}\tau^1}MF_0\eta^{156},
\end{align}
which reproduces the same configuration as the one represented by the disappeared D6-branes.

The same is true for for the other components in Eq.(\ref{hfield}), 
such that $N$ space-filling D6-branes wrapping $\Sigma_{135}+\Sigma_{126}+\Sigma_{123}$ disappear and turn into flux. 

Summarizing our two examples, out of the 20 possible cycles, a bunch of $N$ D6-branes can wrap only 12 cycles in an stable and consistent way, whereas a bunch of $M$ D6-branes can only wrap also 12 cycles. Notice however that in the set of {\it forbidden} eight cycles there are two of them in common. Therefore in the case $N=M$ it is only possible to wrap D6-branes in six three-cycles. 
Details can be consulted in Table 1.

\begin{table}
\begin{center}
\begin{tabular}{||c||c|c||}
\hline\hline
&&\\
D6-branes&$\omega_{(x)}$&${H}_3+\omega J_c$\\
&&\\
\hline\hline
&&\\
Torsion&${\bf N\Sigma_{123}}$&$M\Sigma_{234}$\\
&$N(\Sigma_{134}+\Sigma_{235})$&$M\Sigma_{135}$\\
&$N(\Sigma_{125}+\Sigma_{136})$&$M\Sigma_{126}$\\
&$N(\Sigma_{236}+\Sigma_{124})$&${\bf M\Sigma_{123}}$\\
&&\\
\hline
&&\\
instantonic&${\bf \Sigma_{456}}$&$\Sigma_{156}$\\
&$\Sigma_{146}+\Sigma_{256}$&$\Sigma_{264}$\\
&$\Sigma_{245}+\Sigma_{346}$&$\Sigma_{345}$\\
&$\Sigma_{356}+\Sigma_{145}$&${\bf \Sigma_{456}}$\\
&&\\
\hline\hline
\end{tabular}
\caption{{\small For the twisted six-torus, there are six three-cycles on which D6-branes can be safely wrapped, whereas there are four torsional ones on which $N$ D6-branes become unstable to decay into fluxes and four more on which wrapped D6-branes are FW anomalous. The only option for them is to be instantonic. By turning an extra NS-flux $H_3+\omega J_c$ the same happens for the cycles listed in the third column.}}
\end{center}
\end{table}

\subsection{Transition between fluxes}
The two common cycles, namely $N\Sigma_{123}$ and $\Sigma_{456}$,  open up the possibility to have a chain of topological transformations between different kind of fluxes. To be more specific, let us consider the same case as in example 2 for $N=M$, i.e.,  where the twisted six-torus is threaded by $N$ units of metric and NS- fluxes.

An straightforward observation in that case is the fact that D6-branes are transformed into fluxes by two different instantonic branes: D6 and D8 space-filling branes. The flux which drives the transformation is different for each case. For the instantonic D6 we have a metric flux whereas for the D8-brane is the complexified NS flux $H_3+\omega J_c$. Being the source of the anomaly (in the instantonic brane) different for each case, the remnant fluxes are as well different. However it is possible to show that these two different remnants are indeed connected by a chain of the above two instantonic branes, and in turn establishing a process  which topologically transform one into the other.

To show that, take for instance a RR field strength in the bulk given by $\omega F_2$ such that in terms of the twisted-basis
\begin{align}
-\omega F_2=N\eta^{456}.
\end{align}
Since the form $N\eta^{456}$ is trivial in cohomology, it is represented by also by $N$ space-filling D6-branes wrapping $\Sigma_{123}$. In other words, the transformation between branes and fluxes through the appearance of an instantonic brane works in both directions. The above form is transformed into a bunch of $N$ D6-branes by the appearance of an instantonic anti D6-brane\footnote{The instantonic must be an anti D6-brane for the current to be conserved under $\ov{d}_H$, i.e., $\ov{d}_H J=0$.} wrapping the cycle $\Pi_4$.

Some time later an instantonic D8 appears driving the form $N\eta^{456}$ to be exact in twisted K-theory rather than only in cohomology. The D8-brane turns the $N$ D6-branes into the flux $({H}_3+\omega J_c)\wedge F_0$ as in example 2. In the end we have a process (or an algorithm) which transforms $\omega F_2$ into $({H}_3+\omega J_c) F_0$, i.e., a transition between RR and NS fluxes.

The process can easily been generalized such that
\begin{align}
\omega \ast \ov{F}_{p+2} \leftrightarrow ({H}_3+\omega J_c)\wedge\ast \ov{F}_{p+4},
\end{align}
via the appearance of a D$p$ and a D$(p+2)$ instantonic branes wrapping suitable cycles.

\subsubsection{Example 3: $\widetilde{T}^6/\IZ_2\times\IZ_2$}
Phenomenologically it is more interesting to analyze the $\IZ_2\times\IZ_2$ orbifold of $\widetilde{T}^6$, where the generators of the $\IZ_2$ actions are given by
\begin{align}
\theta_l:&(\eta^i,\eta^{i+3})\rightarrow (-1)^{\kappa_l}(\eta^i,\eta^{i+3})\\
\end{align} 
where
\begin{align}
\kappa_l=
\left\{ \begin{array}{ll}
1 & \textrm{for~}i=2,3\quad\textrm{and~}l=1\\
1& \textrm{for~}i=1,2\quad\textrm{and~}l=2,\\
0 &\textrm{otherwise}.
\end{array}
\right.
\end{align}

The cohomology of invariant forms under the orbifold group $\IZ_2\times\IZ_2$ is also given in Ref.\cite{Marchesano:2006ns}. For us the interesting case comes from the third (co)homology group,
\begin{align}
H_3(\widetilde{T}^6;\IZ)=H^3(\widetilde{T}^6;\IZ)=\IZ^6\times \IZ_N,
\end{align}
where the only one $N$-torsion form is
\begin{align}
N\eta^{456}=d(\eta^{14}+\eta^{25}+\eta^{36}),
\end{align}
and the non-closed form is $\eta^{123}$. Hence, out of the 8 possible 3-cycles of $\widetilde{T}^6/\IZ_2\times\IZ_2$ there are only 6 cycles on which $N$ D6-branes can be wrapped under the presence of metric fluxes as shown in \cite{Marchesano:2006ns}. Hence $N$ D6-branes wrapping $\Sigma_{123}$ are non-stable to decay into $\omega F_2$.

By turning on $M$ units of NS-flux ${H}_3+\omega J_c$ the number of ``forbidden'' cycles increases since now there are six- more cycles on which $M$ D6-branes are either unstable to decay into fluxes or inconsistent for being FW anomalous\footnote{This is they can play the role of instantonic branes for a bunch of D0-branes in the twisted six-torus. This would trigger a cascade in the gauge theory of type IIB D3-branes sitting on a point in the twisted torus.} as listed in Table 2.

\begin{table}
\begin{center}
\begin{tabular}{||c||c|c||}
\hline\hline
&&\\
D6-branes&$\omega_{(x)}$&$\ov{H}_3$\\
&&\\
\hline\hline
&&\\
Torsion&${\bf N\Sigma_{123}}$&$M\Sigma_{234}$\\
&&$M\Sigma_{135}$\\
&&$M\Sigma_{126}$\\
&&${\bf M\Sigma_{123}}$\\
&&\\
\hline
&&\\
instantonic&${\bf \Sigma_{456}}$&$\Sigma_{156}$\\
&&$\Sigma_{264}$\\
&&$\Sigma_{345}$\\
&&${\bf \Sigma_{456}}$\\
&&\\
\hline\hline
\end{tabular}
\caption{{\small For the $\IZ_2\times\IZ_2$-orbifold twisted six-torus there are 6 three-cycles on which D6-branes can be safely wrapped, whereas there is a torsional one on which $N$ D6-branes become unstable to decay into fluxes and another one which is completely forbidden since it is FW anomalous. However for the same torus threaded by a NS-flux there are not safe cycles in the sense that the only 4-cycles on which space-filling D6-branes can be wrapped are torsional, driving the branes to be unstable. }} 
\end{center}
\end{table}

It is interesting to notice that in this case, D6-branes can wrap only 4 three-cycles which turn to be torsional. Hence all possible consistent D6-branes are unstable to decay into a configuration of fluxes $F_0({H}_3+\omega J_c)$. The only way to avoid this situation is canceling trivially the FW anomaly by vanishing the extra flux,
\begin{align}
H_3 + \omega J_c=0
\end{align}
as in \cite{Camara:2005dc, Villadoro:2005cu, Villadoro:2006ia}.
It would be interesting to study the complete cohomology group for the orbifolded twisted torus which takes into account the twisted forms as well. It is possible that in such case there are cycles neither torsional nor anomalous.

\section{Conclusions}

There are global effects on the stability of D-branes in a compactification setup threaded by external fluxes. Since every D-brane wrapping a cycle which supports non-trivial units of NS-NS flux is potentially Freed-Witten anomalous, it is required either to cancel the external flux on the worldvolume of D-branes, or to add other branes to cancel the anomaly. Usually phenomenological models rather consider the first case in order to avoid extra branes. 

However there are D-branes which although are FW anomaly-free, can be nevertheless unstable.
This is physically interpreted as a topological transformation between D-branes and fluxes through the appearance of instantonic branes as it was shown in the seminal paper by Maldacena, Moore and Seiberg (MMS)\cite{Maldacena:2001xj}. Hence, taking only D-branes on which the extra flux vanishes is not always the solution to avoid the appearance of topological effects as the above mentioned transition. Formally such transition is described by a mathematical algorithm known as the Atiyah-Hirzebruch Spectral Sequence which roughly speaking, connects cohomology with twisted K-theory.\\

So, in this paper we have studied such effects in the twisted six-torus (particularly in type IIA) realized by taking T-duality on the flat torus threaded by extra NS-flux (in type IIB). Following Ref.\cite{Marchesano:2006ns} where the author shows that all NS-flux effects on the stability of D-branes are mapped into the torsion component of the twisted six-torus cohomology, we proposed a T-dual version of the AHSS. 

Whereas for the flat six-torus the effects of NS-flux was encoded in the second step of the AHSS, in the twisted one they are encoded in the torsion part of cohomology, i.e., in the first step. According to the physical interpretation given by MMS, we found that $N$ D$p$-branes wrapping a torsion cycle will decay into fluxes by encountering an instantonic D$p$-brane wrapped in a $(p+1)$-cycle supporting the metric fluxes. The remnant fluxes are given by the coupling between metric fluxes and the magnetic field strength related to the instantonic brane. This allows to transform in a topological sense, D$p$-branes into $-\omega\ast F_{p+2}$ which is in accordance to the the fact that
\begin{align}
d\ast F_{p+1}=-\omega\ast F_{p+1}.
\end{align}
We also studied the equations of motion for RR potential in a background threaded by metric fluxes as well as the field theory content on the worldvolume of a D$p$-brane supporting metric fluxes. The results of these systems together with a particular example, are in agreement with the proposed T-dual version of the AHSS. 

However, although the second step in the AHSS, denoted as $E_3^p(T^6)$ and  involving the extra NS-flux,  is the final one to connect cohomology classes with K-theory ones (in the absence of fixed points), the T-dual version of it cannot finish at the first step $E_1^p(\widetilde{T}^6)$ (T-dual of $E_3^p(T^6)$) since the refinement has taken place only at the torsion part of cohomology. We have still to refine the non-torsion part of cohomology. Hence by turning on an extra complex three-form, composed of the NS-flux and the non-closed K\"ahler form we proposed that the second step in the T-dual AHSS is to take forms which are closed but non-exact under the differential map
\begin{align}
\ov{d}_3= \wedge (H_3+\omega J).
\end{align}
The group of such forms is denoted by $\ov{E}^p_3(\widetilde{T}^6)$. The correspondent physical interpretation is that now those branes representing trivial forms under $\ov{d}_3$ would decay into a configuration of fluxes by encountering an instantonic D$(p+2)$-brane wrapped in a cycle which in turn supports the complex flux $H_3+\omega J$. The remnant is the coupling between the complex three-form and the magnetic field strength for the instantonic brane. Hence D$p$-branes, representing non-trivial forms in $\ov{E}_i^p(\widetilde{T}^6)$ are topologically transformed into $(H_3+\omega J)\wedge\ast F_{p+4}$. We also analyze the correspondent D-brane action threaded by these external fluxes and the tadpole generated in the worldvolume of a D-brane supporting such extra flux. The results are as well in agreement with the proposed second step in the T-dual AHSS. A particular example shows the validity of our proposal.

Interesting enough, we found that in some cases, as for three-cycles in our examples, the groups  $\ov{E}_1^3(\widetilde{T}^6)$ and $\ov{E}_3^3(\widetilde{T}^6)$ share some exact three-forms. This allows to construct a chain of instantonic branes in which the flux $\omega \ast F_{p+2}$ is transformed into $(H_3+\omega J)\wedge\ast F_{p+4}$, relating NS-NS fluxes with RR ones and viceversa. An explicit example is also provided, showing that for a type IIA compactification on a twisted six-torus threaded by a metric flux in coordinates 456 and a NS-flux compatible with supersymmetric conditions, there are only 6 cycles out of 20, where a space-filling D6-brane can be wrapped.

Moreover by studying the case of the twisted six-torus orbifolded by $\IZ_2\times\IZ_2$, we find that all D6-branes wrapping the invariant (untwisted) cycles, are either unstable to decay into fluxes or anomalous in the context of Freed-Witten. In cases like that, phenomenological models are not protected for instabilities unless $H_3+\omega J=0$ as in \cite{Camara:2005dc, Villadoro:2005cu}. It is important to say that we have consider only forms and cycles invariant under the action of the above discrete group. It would be interesting to extend our study to twisted forms as well.

We hope that these effects on the stability and transformation between branes and fluxes, could help to provide with new constraints to phenomenological models where the presence of both, branes and fluxes are fundamental.\\

\bigskip

\begin{center}
{\bf Acknowledgments}
\end{center}

It is a pleasure to thank Prof. Hans-Peter Nilles for great support during my stay at Bonn University. My gratitude to Kang-Sin Choi, Jarah Evslin,  Hugo Garc\'{\i}a-Compe\'an, Amihay Hanany, Andrei Micu, Hans-Peter Nilles, Kin-Ya Oda and Angel Uranga for enlighting discussions at different parts of this project. Especially, I would like to thank Gianmassiamo Tasinato for great discussions and collaboration. His ideas are fundamental part of this work. Thanks to the RIKEN laboratory of theoretical physics for kind hospitality in which part of this research was done. I am thankful to  Hugo Garc\'{\i}a-Compe\'an for great support in my arriving at CINVESTAV-Monterrey. This work was partially supported by European Union 6th framework program MRTN-CT-2004-503369 ``Quest for unification'' and MRTN-CT-005104 ``Forces Universe'' in Bonn and by CONACYT-Mexico under the program ``repatriaci\'on''.

\appendix

\section{The Atiyah-Hirzebruch Spectral Sequence}

Let us give a brief review on the Atiyah-Hirzebruch Spectral Sequence (AHSS) (for further analysis, see \cite{Diaconescu:2000wy, Maldacena:2001xj, Evslin:2006cj}). Essentially it is an algorithm which relates integral cohomology with twisted K-theory. The relation involves the construction, by a finite number of steps, of twisted K-theory classes from integral cohomology classes.  In general, an integral cohomology class $[\omega_p]\in H^p(X;\IZ)$ does not come from a twisted K-theory class $[x]\in K(X)$. Hence, the  algorithm begins with cohomology. At this step, cohomology is the first approximation to twisted K-theory and it is denoted by $E_1(X)$.\\ 

At the $p$th step, the approximate group is denoted by $E_p(X)$, where
\begin{align}
K(X)\sim E^p_m \equiv \frac{\text{Ker~} d_m|_{E^p_{m-2}}}{\text{Im~ }d_m|_{E^{p-m}_{m-2}}}\;,
\end{align}
such that, the first approximation is
\begin{align}
E_1(X) = \oplus_p H^p(X;\IZ)\;.
\end{align}

The second step is to consider forms which are closed under the differential map
\begin{align}
d^3 \equiv \wedge H_3\;,
\end{align}
with $d^3:H^p(Z;\IZ) \rightarrow H^{p+3}(X;\IZ)$, but discarding those which are exact. This defines the group
\begin{align}
E^p_3(X)=\frac{\text{Ker~}d_3|_{H^p}}{\text{Im~ }d_3|_{H^{p-3}}}\;.
\end{align}

After this step, only those forms which are closed will survive and represent stable D-branes in string theory provided the NS-NS field is identified with $H_3$. Those forms which are not closed represents the instantonic branes we have discussed in this paper. Finally, $(p+3)$-forms which belong to the trivial class satisfy
\begin{align}
d^3 \omega_p ~=~\omega\wedge H_3~=~\sigma_{p+3}\;,
\end{align}
and represent branes which can nevertheless be unstable. Notice that, according to the integral class of $H_3$ (upon the isomorphism with the field), such forms belong to torsion classes in K-theory.\\

One can go further, defining several groups in order to get a closer approximation to K-theory. However, the algorithm ends after a finite number of steps. At the end, one gets a group which is called ``associated graded group'' $Gr(X)$ given by,
\begin{align}
Gr(K_H(X))=\oplus_p E^p(X)\;.
\end{align}

This group, is in some cases\footnote{In the context of string theory, this happens in the absence of orientifold planes.} the K-theory group. However, in other cases it is necessary to solve an extension problem, since
\begin{align}
Gr(K_H(X))=\oplus_p K_{H,p}(X)/ K_{H, p+1}(X)\;.
\end{align}

In cases as in this paper, the second step is the final approximation.

\bibliography{ttori}

\providecommand{\bysame}{\leavevmode\hbox to3em{\hrulefill}\thinspace}
\begin{thebibliography}{10}

\bibitem{Grana:2005jc}
M.~Grana, \emph{Flux compactifications in string theory: A comprehensive
  review}, Phys. Rept. \textbf{423} (2006), 91--158,  \texttt{hep-th/0509003}.

\bibitem{Scherk:1979zr}
J.~Scherk and J.~H. Schwarz, \emph{How to get mass from extra dimensions},
  Nucl. Phys. \textbf{B153} (1979), 61--88.

\bibitem{Kaloper:1999yr}
N.~Kaloper and R.~C. Myers, \emph{The o(dd) story of massive supergravity},
  JHEP \textbf{05} (1999), 010,  \texttt{hep-th/9901045}.

\bibitem{Freed:1999vc}
D.~S. Freed and E.~Witten, \emph{{Anomalies in string theory with D-branes}},
  (1999),  \texttt{hep-th/9907189}.

\bibitem{Hanany:1996ie}
A.~Hanany and E.~Witten, \emph{{Type IIB superstrings, BPS monopoles, and
  three-dimensional gauge dynamics}}, Nucl. Phys. \textbf{B492} (1997),
  152--190,  \texttt{hep-th/9611230}.

\bibitem{Witten:1998xy}
E.~Witten, \emph{Baryons and branes in anti de sitter space}, JHEP \textbf{07}
  (1998), 006,  \texttt{hep-th/9805112}.

\bibitem{Maldacena:2001xj}
J.~M. Maldacena, G.~W. Moore, and N.~Seiberg, \emph{{D-brane instantons and
  K-theory charges}}, JHEP \textbf{11} (2001), 062,  \texttt{hep-th/0108100}.

\bibitem{Diaconescu:2000wy}
D.-E. Diaconescu, G.~W. Moore, and E.~Witten, \emph{{E(8) gauge theory, and a
  derivation of K-theory from M- theory}}, Adv. Theor. Math. Phys. \textbf{6}
  (2003), 1031--1134,  \texttt{hep-th/0005090}.

\bibitem{Evslin:2006cj}
J.~Evslin, \emph{{What does(n't) K-theory classify?}},  (2006),
  \texttt{hep-th/0610328}.

\bibitem{Evslin:2006tc}
J.~Evslin and H.~Sati, \emph{Can d-branes wrap nonrepresentable cycles?}, JHEP
  \textbf{10} (2006), 050,  \texttt{hep-th/0607045}.

\bibitem{Minasian:1997mm}
R.~Minasian and G.~W. Moore, \emph{{K-theory and Ramond-Ramond charge}}, JHEP
  \textbf{11} (1997), 002,  \texttt{hep-th/9710230}.

\bibitem{Witten:1998cd}
E.~Witten, \emph{{D-branes and K-theory}}, JHEP \textbf{12} (1998), 019,
  \texttt{hep-th/9810188}.

\bibitem{Schwarz:1999vu}
J.~H. Schwarz, \emph{{TASI lectures on non-BPS D-brane systems}},  (1999),
  \texttt{hep-th/9908144}.

\bibitem{Witten:2000cn}
E.~Witten, \emph{{Overview of K-theory applied to strings}}, Int. J. Mod. Phys.
  \textbf{A16} (2001), 693--706,  \texttt{hep-th/0007175}.

\bibitem{Moore:2003vf}
G.~W. Moore, \emph{K-theory from a physical perspective},  (2003),
  \texttt{hep-th/0304018}.

\bibitem{Collinucci:2006ug}
A.~Collinucci and J.~Evslin, \emph{Twisted homology},  (2006),
  \texttt{hep-th/0611218}.

\bibitem{Uranga:2002vk}
A.~M. Uranga, \emph{D-brane, fluxes and chirality}, JHEP \textbf{04} (2002),
  016,  \texttt{hep-th/0201221}.

\bibitem{Townsend:1996em}
P.~K. Townsend, \emph{Brane surgery}, Nucl. Phys. Proc. Suppl. \textbf{58}
  (1997), 163--175,  \texttt{hep-th/9609217}.

\bibitem{Evslin:2001cj}
J.~Evslin and U.~Varadarajan, \emph{{K-theory and S-duality: Starting over from
  square 3}}, JHEP \textbf{03} (2003), 026,  \texttt{hep-th/0112084}.

\bibitem{Bekaert:2002cz}
X.~Bekaert, \emph{Issues in electric-magnetic duality},  (2002),
  \texttt{hep-th/0209169}.

\bibitem{deAlwis:2006cb}
S.~P. de~Alwis, \emph{Transitions between flux vacua},  (2006),
  \texttt{hep-th/0605184}.

\bibitem{Bouwknegt:2003vb}
P.~Bouwknegt, J.~Evslin, and V.~Mathai, \emph{{T-duality: Topology change from
  H-flux}}, Commun. Math. Phys. \textbf{249} (2004), 383--415,
  \texttt{hep-th/0306062}.

\bibitem{Bouwknegt:2003wp}
P.~Bouwknegt, J.~Evslin, and V.~Mathai, \emph{{On the topology and H-flux of
  T-dual manifolds}}, Phys. Rev. Lett. \textbf{92} (2004), 181601,
  \texttt{hep-th/0312052}.

\bibitem{Marchesano:2006ns}
F.~Marchesano, \emph{D6-branes and torsion}, JHEP \textbf{05} (2006), 019,
  \texttt{hep-th/0603210}.

\bibitem{Cascales:2003zp}
J.~F.~G. Cascales and A.~M. Uranga, \emph{{Chiral 4d N = 1 string vacua with
  D-branes and NSNS and RR fluxes}}, JHEP \textbf{05} (2003), 011,
  \texttt{hep-th/0303024}.

\bibitem{Camara:2005dc}
P.~G. Camara, A.~Font, and L.~E. Ibanez, \emph{{Fluxes, moduli fixing and
  MSSM-like vacua in a simple IIA orientifold}}, JHEP \textbf{09} (2005), 013,
  \texttt{hep-th/0506066}.

\bibitem{Villadoro:2005cu}
G.~Villadoro and F.~Zwirner, \emph{{N = 1 effective potential from dual
  type-IIA D6/O6 orientifolds with general fluxes}}, JHEP \textbf{06} (2005),
  047,  \texttt{hep-th/0503169}.

\bibitem{Villadoro:2006ia}
G.~Villadoro and F.~Zwirner, \emph{D terms from d-branes, gauge invariance and
  moduli stabilization in flux compactifications}, JHEP \textbf{03} (2006),
  087,  \texttt{hep-th/0602120}.

\bibitem{Kachru:2002sk}
S.~Kachru, M.~B. Schulz, P.~K. Tripathy, and S.~P. Trivedi, \emph{New
  supersymmetric string compactifications}, JHEP \textbf{03} (2003), 061,
  \texttt{hep-th/0211182}.

\bibitem{Taylor:2000za}
W.~Taylor, \emph{{D2-branes in B fields}}, JHEP \textbf{07} (2000), 039,
  \texttt{hep-th/0004141}.

\end{thebibliography}
\addcontentsline{toc}{section}{Bibliography}
\bibliographystyle{TitleAndArxiv} 
\end{document}